# Phase-dependent Andreev molecules and superconducting gap closing in coherently coupled Josephson junctions

Sadashige Matsuo[1], Takaya Imoto[1,2], Tomohiro Yokoyama[3], Yosuke Sato[1], Tyler Lindemann[4,5], Sergei Gronin[4], Geoffrey C. Gardner[4], Sho Nakosai[6], Yukio Tanaka[6], Michael J. Manfra[4,5,7,8], Seigo Tarucha[1,9]

[1] *Center for Emergent Matter Science, RIKEN, Saitama 351-0198, Japan*
[2] *Department of Physics, Tokyo University of Science, Tokyo 162-8601, Japan*
[3] *Department of Materials Engineering Science, Osaka University, Osaka 560-8531, Japan*
[4] *Birck Nanotechnology Center, Purdue University, West Lafayette, Indiana 47907, USA*
[5] *Department of Physics and Astronomy, Purdue University, West Lafayette, Indiana 47907, USA*
[6] *Department of Applied Physics, Nagoya University, Nagoya 464-8603, Japan*
[7] *School of Materials Engineering, Purdue University, West Lafayette, Indiana 47907, USA*
[8] *Elmore Family School of Electrical and Computer Engineering, Purdue University, West Lafayette, Indiana 47907, USA*
[9] *RIKEN Center for Quantum Computing, RIKEN, Saitama 351-0198, Japan*

sadashige.matsuo@riken.jp
tarucha@riken.jp

## Abstract

The Josephson junction (JJ) is an essential element of superconducting (SC) devices for both fundamental and applied physics. The short-range coherent coupling of two adjacent JJs forms the Andreev molecule states (AMSs), which will provide a new ingredient to engineer the SC transport in JJs and control the Andreev qubits. However, no experimental evidence of the AMSs in the coupled JJs has been reported. Here we provide the tunnel spectroscopic results of electrically controllable two planar JJs sharing one SC electrode. We discover that the coupled JJ results are highly modulated from the single JJ results, due to formation of the phase-dependent AMSs, meaning that the two JJs are coherently coupled. In addition, the superconducting gap closing due to the AMS formation is observed. Our results would help in understanding the microscopic mechanism of the coherent coupling and promoting the AMS physics to apply for research of the topological superconductivity and quantum information technology.

The Josephson junction (JJ) is a representative superconducting (SC) device consisting of two weakly-linked superconductors through insulators or normal conductors [1]. The JJs have been studied to engineer quantum effects in solid-state devices, enabling to realize SC quantum computing and highly sensitive magnetic sensors. The recent development of JJs on the superconductor-semiconductor heterostructures has provided platforms for more exotic physics such as Andreev (spin) qubits [2–6], SC qubits with gate tunability [7,8] or Majorana zero modes (MZMs) [9–12]. From these aspects, engineering of coherent coupling between two JJs is an essential ingredient to explore novel SC phenomena, establish new control methods of JJs, and manage the qubit operation.

Recently a concept of short-range coherent coupling between two JJs has been proposed as Andreev molecule states (AMSs) [13–15]. In a single JJ consisting of two superconductors and a semiconductor, electrons are confined by the Andreev reflection at interfaces, forming the Andreev bound states (ABSs) [16–18]. For example, in the case of two adjacent JJs sharing one SC electrode, the ABS wavefunctions in the different JJs penetrate to the shared SC and overlap, which produces the coherently coupled wavefunctions called as AMSs. This is an analogy of the molecular orbital states formed by the coherent coupling of two atomic states. The recent experimental efforts on the coherently coupled JJs have revealed nonlocal SC transport assigned to the AMSs such as the nonlocal Josephson effect [19]. Additionally, the AMS physics in the coupled JJs may provide novel insights on the SC transport intermediated by the Cooper pair splitting in a parallel double quantum dot or double nanowire JJ, which can be regarded as the two JJs sharing two SC electrodes [20–23]. In order to understand the microscopic mechanisms of such SC transport for the novel SC device functionality and engineer operation and integration of the Andreev qubits using the coupled JJs [24], observation of the AMSs in the coupled JJs is indispensable. However, experimental evidence of the AMS formation in the coupled JJs is lacking although there are experimental reports of the AMS signatures formed in SC junctions other than JJs [25–27].

Additionally, the AMSs in the coupled JJ structures contribute to and develop the topological physics in the JJs. In the case of conventional two-terminal JJs, the Kramers degeneracy, which demands the energy levels of half-integer spin systems, is doubly degenerate in the presence of time-reversal symmetry even with the spin-rotation symmetry broken by the spin-orbit interactions (SOI), leads to a small amount of SC gap easily caused. For example, such SC gap in the short ballistic JJs closes only when the junction transmission is unity [16]. Lifting this Kramers degeneracy by breaking the time-reversal symmetry in the JJs with SOI leads to various exotic SC phenomena including the realization of MZMs. Multiterminal JJ structures which have more than three or more SC electrodes contacted on a single normal metal [25,28–38] enable to break the time-reversal symmetry only by controlling the phase differences with no strong magnetic fields. As a signature of the Kramers degeneracy lifting in the multiterminal JJs, the SC gap closing only by the phase control is predicted [28,37,38] and related experimental results have been reported [39]. Observation of this SC gap closing is an important step to realize and control more exotic SC phenomena predicted in the multiterminal JJs including Weyl fermion physics [29,31,40,41]. In the multiterminal JJ case, the ABSs are formed in the single normal metal, but in the coupled JJ case, the AMSs are not restricted to the single normal metal and spread over the two JJs. Therefore, the coupled JJ structure enables to design more diverse physics of AMSs hosted in various combinations of different JJs.

Here we experimentally study the Andreev spectrum in single and coupled JJs by tunnel spectroscopy [11,42] to elucidate the phase-dependent AMSs. For this sake, we fabricate an SC device of two JJs named JJL and JJR sharing an SC electrode from a high-quality InAs quantum well covered with the epitaxially grown thin Aluminum layer [43–45]. The scanning electron microscopic image and the schematic image of the coupled JJ device are shown in Figs. 1(a) and (b), respectively. The separation between the two JJs corresponding to the width of the shared SC electrode is designed as 150 nm which is sufficiently shorter than the coherence length of Aluminum (Al). The junction length and width are 80 nm and 1.9 μm, respectively. The two JJs are each respectively embedded in a SC loop which encloses the same area, to induce the same phase difference in JJL and JJR. With definition of $\phi_L$ and $\phi_R$ of the phase differences on JJL and JJR as $\phi_L = \phi_s - \phi_l$ and $\phi_R = \phi_r - \phi_s$, respectively, the out-of-plane magnetic field $B$ changes the phase differences with $\phi_L = \phi_R$. Here $\phi_l$, $\phi_s$, and $\phi_r$ represent the phases of the left, shared, and right SC electrodes, respectively. The main concept of this design is to compare the spectroscopic results in the single JJ and the coupled JJ cases using the same device but by electrically independently controlling the two JJs. For this sake, we have fabricated the gate electrodes as highlighted in yellow in Fig. 1 (a). The gate electrodes on JJL and JJR are used to control the planar JJs with the gate voltages of $V_{gL}$ and $V_{gR}$, respectively. Additionally, three gate electrodes are prepared to electrically form the quantum point contacts on the edges of the planar JJs (see supplementary Note 2). The gate voltages $V_{g1}, V_{g2}$, and $V_{g3}$ are applied on the electrodes as depicted in Fig. 1(b). We name the point contact formed on the edge of JJL (JJR) as QPCL (QPCR). When performing the spectroscopy of JJL, we pinch off the QPCR and detect the tunnel current through the QPCL as the schematic circuit diagram in Fig. 1(b). We measure this device at 10 mK of the base temperature in our dilution refrigerator. For the tunnel spectroscopy, we apply a D.C. bias voltage $V$ with a small oscillation component (5 μ$V$) and measure the tunnel currents through QPCL and QPCR by Lock-in amplifiers. Then we obtain $G_L$ and $G_R$ of differential conductance through the QPCL and QPCR, respectively.

First, we perform the tunnel spectroscopy of the single JJs. For the sake, we measure $G_L$ with JJR pinched off $(V_{gL}, V_{gR})$= (0 V, -6 V). In the measurement, QPCR is pinched off also and QPCL is tuned to allow $G_L \sim 0.10 e^2/h$. Then the obtained $G_L$ as a function of $V$ and $B$ is indicated in Fig. 2(a). As clearly seen, the subgap structure emerges inside the Aluminum SC gap energy (~0.18 meV). The observed SC gap inside the Aluminum gap energy is modulated periodically as a function of $B$. The period is around 0.126 mT consistent with the expected period evaluated from the loop area (0.172 mT). This periodic modulation has been reported in previous studies [11,18,42] and assigned to the phase modulation of ABSs in the single ballistic JJ [16,42]. Therefore, the oscillation period of the gap is equivalent to $2\pi$ and $B$ producing the maximum (minimum) gap corresponds to $\phi_L = 0 \ (\text{mod} \ 2\pi)$ ($\phi_L = \pi \ (\text{mod} \ \pi)$). It is noted that JJR is pinched off so that $\phi_R$ is not considered here (see supplementary note 3 and 4, and Figs. S3 and S5). In our results, the maximum (minimum) induced gap is around 0.1 meV (0.08 meV). The induced SC gap is defined as the ABSs with no momentum along the junction and the ABSs with finite momentum fill the states outer of the gap. These results imply that our method of tunnel spectroscopy using QPCL correctly detects the Andreev spectrum in JJL. We note that the jump of the data at $B = 0.066$ mT occurs because of the charge jump around QPCL.

Then, we move to the characterization of JJR with $(V_{gL}, V_{gR}) = $ (-6 V, 0 V). We measure $G_R$ as a function of $V$ and $B$ with QPCL and JJL pinched off. The results are shown in Fig. 2(b). As well as the JJL results, the periodic oscillation of the SC gap is observed, which is the ABS oscillation of the single JJR as a function of $\phi_R$. In the JJR results, the maximum (minimum) induced gap is around 0.12 meV (0.075 meV). Comparing the JJR result with the JJL result, the oscillation periods in both results are consistent, which reflects the two SC loops have the same loop area. Therefore, the results imply that our device works correctly to detect the ABSs of both the single planar JJs with the phase biased by the out-of-plane magnetic field.

Next, we explore the tunnel spectroscopy of JJL with JJR on $(V_{gL}, V_{gR}) = (0\ \text{V}, 0\ \text{V})$. Here two JJs are turned on, then both of $\phi_L$ and $\phi_R$ evolve as a function of $B$ with $\phi_L = \phi_R$ assured by the same SC gap oscillation period of the single JJL and JJR results. The obtained spectroscopic result ($G_L$ as a function of $B$ and $V$) is shown in Fig. 2(c). Compared with the single JJ result in Fig. 2(a), the Andreev spectrum is drastically modulated. This drastic change is invoked by turning on JJR, namely turning on the coherent coupling between two JJs. Therefore, the observed spectroscopic result and the change from the single JJ results are assigned to the formation of AMSs in the coupled JJs. Here a significant feature that cannot be found in the single JJ results is observed. In the coupled JJ case, the SC gap becomes maximal where the gap in the single JJL result in Fig. 2(a) produces the minimum. This means that $\phi_L = \phi_R = 0, \pi \pmod{2\pi}$ give the maximal SC gap in Fig. 2(c) and the minimum SC gap in Fig. 2(c) is realized in $0 < \phi_L = \phi_R < \pi$.

When the coherent coupling between JJL and JJR modulates the JJL Andreev spectrum, the JJR spectrum should also be modulated as well. Then we confirm the JJR spectrum by the tunnel spectroscopy of JJR with $(V_{gL}, V_{gR}) = (0\ \text{V}, 0\ \text{V})$. The obtained $G_R$ as a function of $B$ and $V$ is shown in Fig. 2(d). As explicitly found, the same feature as in the JJL results appears. The consistency obtained between the JJL result in Fig. 2(c) and the JJR result in Fig. 2(d) assures that the Andreev spectrum modulation is induced by the coherent coupling between the two planar JJs. We note that these features are reproducible (see supplementary note 1 and Figs. S1 and S2). Particularly, both results exhibit the same dependence on the phase differences, which means that the coherent coupling maintains the phase coherence.

To evaluate the minimum SC gap energies in the coherent coupling case, we exhibit the line profiles at $B = 0.230, 0.273$, and $0.295$ mT, in Figs.2(a)-(d) in red, green, and blue in Figs. 2(e)-(h), respectively. Note that the SC gap changes from the maximum to the minimum in this $B$ field range in the single JJ, while from the maximum to the minimum and then to the maximum in the coupled JJs.

Figs. 2(e) and (f) indicate the line profiles of the single JJL and JJR results in Figs. 2(a) and (b), respectively. As seen in Figs. 2(e) and (f), the Andreev spectrum is always gapped. This holds at any $B$ field. On the other hand, in Figs. 2(g) and (h), the green line profile does not touch the $G = 0$ line. Therefore, the SC gap looks closed in our setup resolution (~0.01 mV). Note that the same result is obtained for each SC gap minimum.

To reveal that the modulated Andreev spectrum reflects the AMS properties owing to the coherent coupling through the shared SC electrode, we perform numerical calculations on the tight-binding model (see supplementary note 7 and Fig. S7). Figures 3 (a) and (b) indicate the numerical calculation results without and with Rashba SOI, respectively. The numerical calculation results explain the properties of the experimentally observed spectrum. For example, the SC gap becomes minimal between $0 < \phi_L = \phi_R < \pi$ and then local maximal at $\phi_L = \phi_R = 0, \pi$. This behavior is caused by a coherent coupling between the ABSs in two JJs, which means the formation of the AMSs. Furthermore, Fig. 3 (c) is an enlarged view around the minimal SC gap in Fig. 3(b), indicating the SC gap is closed in the AMS spectrum owing to the SOI. Figure 3 (d) exhibits the energy of the positive lowest Andreev state as a function of $\phi_L$ and $\phi_R$. At the mean free path being 217 nm roughly consistent with that in our InAs quantum well, we obtain the gap closing in the AMS spectrum in 19% of samples in the simulation (see supplementary note 8 and Fig. S8).

In the coupled JJ devices, realization of topological SC states only by the phase control is theoretically predicted [46]. In the prediction, the topological SC state is formed and the MZMs appear between the two SC gap closing points as seen in Fig. 3(c). For this realization, it is essential that the quantum well holds different Fermi velocities for the spin-up and spin-down states. In our device, this might be induced by mixing of the subbands formed by the confinement [47–49]. When the velocity difference is small, a space of the two SC gap closing points could be tiny in the numerical calculation and then the SC gap closing would be observed to emerge at a single point (see supplementary notes 7 and 8 and Figs. S7 and S8). In this scenario, our finding of the SC gap closing might be related to the topological SC states in the coupled JJ systems.

A similar SC gap closing is theoretically predicted in multiterminal JJs [28,37] and the related experimental results have been reported [39]. In theory, it is revealed that the SC gap closing is allowed when the normal metal of the multiterminal JJs holds the strong SOIs and the three SC phases encircle the origin, meaning $\pi/2 \leq \phi_L = \phi_R \leq 3\pi/2$ in our definition. This reflects that the SC gap closing occurs when the Kramers degeneracy is lifted by breaking the time-reversal symmetry in the JJs with the broken spin-rotation symmetry. It is noted that the SC gap closing (lifting the Kramers degeneracy) in JJs is one of the necessary ingredients to realize the exotic SC phenomena originating from the single Fermion states such as Majorana zero modes, Weyl singularity, and Andreev spin qubits [29,31,39]. In the multiterminal JJs, the time-reversal symmetry can be broken only by controlling the SC phases. Therefore, this SC gap closing is an important signature of breaking the Kramers degeneracy in the junctions holding the SOIs. This resembles our situation in that the AMSs consist of the ABSs in the different JJs coupled by the multipair processes of the double elastic cotunneling or double crossed Andreev reflection [13] and the normal metal of InAs quantum well holds the strong SOI. Therefore, we can conclude that the observed SC gap closing indicates that the Kramers degeneracy in the Andreev spectrum including the SOI can be lifted by controlling the phase differences in the coupled JJ devices. This leads to the use of the coupled JJs for the SC phenomena predicted in the multiterminal JJs [29,31,39] and design more exotic phenomena using the coupled JJs in which the respective JJs consists of the different normal metals.

Subsequently, we explore the gate voltage dependence of the AMSs and the SC gap closing. Especially, we focus on the nonlocal gate control effect on the AMS results. Figure 4(a) shows $G_R$ as a function of $B$ and $V$ at $V_{gL}$=-1, -2, -3, -4, and -5 V with $V_{gR}$= 0 V. It is clear that the AMS features disappear and the ABSs in the single JJR emerge for $V_{gL} \leq$-3 V. Then we fix $B$=-0.03 mT at the SC gap closing and sweep $V_{gL}$ as shown in Fig. 4(b). The gap-closing behavior appears robust for $V_{gL}$ from 0 to $-2$ V and gapped for $V_{gL} \lesssim -2$ V. With Fig. 4(a), this indicates that JJL is pinched off around $V_{gL}$=-2 V and the SC gap closing disappears when the AMSs disappear. This is supported by the magnetic field dependence of $G_R$ measured as a function of $V_{gL}$ shown in Fig. 4(c). The two bright vertical lines are observed in the gate voltage range of -2 V $< V_{gL} \leq$ 0 V. This indicates the SC gap closing of JJR occurs in -2 V $< V_{gL} \leq$ 0 V and almost at the same magnetic fields in the nonlocal gate control. Similar behavior is also observed in the nonlocal gate control of the JJL (see supplementary note 5 and Figs. S4 and S5). Furthermore, we confirm that the SC gap closing is robust against the control of $V_{gR}$ as long as the AMSs are formed (see supplementary note 6 and Fig. S6). These results reveal that the AMSs are formed when both JJL and JJR exist and are coherently coupled with each other even if their carrier densities are different. In addition, the SC gap closing is robust against the control of the carrier densities of JJL and JJR. On the other hand, the numerical calculation does not always produce the SC gap closing in the spectrum. This difference may be related to a hidden physical mechanism in the coherently coupled JJs or the finite energy resolution of the tunnel spectroscopy of our setup. Further studies are demanded. We note that similar spectroscopic results of a three-terminal JJ on an InAs quantum well have been reported on arXiv [50] while preparing this manuscript.

As a summary, we perform the tunnel spectroscopy of the electrically controllable planar JJs embedded in the two SC loops. We find that the Andreev spectrum of single JJs indicates the periodic ABS oscillation as expected in ballistic JJs while the coupled JJs exhibit the strongly modulated structures holding the SC gap closing. The modulated Andreev spectrum is discovered in both of the two JJs and reproduced by the theoretical calculation. From these results, it is concluded that the AMSs in the coupled planar JJs are detected. Furthermore, it is discovered that the SC gap closing resembles the theoretical prediction in multi-terminal JJs. This study will contribute to elucidating the microscopic mechanisms of the AMS formation and developing quantum control and integration of the Andreev qubits. Additionally, our results suggest that the coupled planar JJs will provide a new platform to engineer Majorana zero modes and Weyl singularities realized by the multiple phase differences.

# Method

## Sample growth

The wafer structure has been grown on a semi-insulating InP substrate by molecular beam epitaxy. The stack materials from the bottom to top are a 100 nm $In_{0.52}Al_{0.48}As$ buffer, a 5 period 2.5 nm $In_{0.53}Ga_{0.47}As$/2.5 nm $In_{0.52}Al_{0.48}As$ superlattice, a 1μm thick metamorphic graded buffer stepped from $In_{0.52}Al_{0.48}As$ to $In_{0.84}Al_{0.16}As$, a 33 nm graded $In_{0.84}Al_{0.16}As$ to $In_{0.81}Al_{0.19}As$ layer, a 25 nm $In_{0.81}Al_{0.19}As$ layer, a 4 nm $In_{0.81}Ga_{0.19}As$ lower barrier, a 5 nm InAs quantum well, a 10 nm $In_{0.81}Ga_{0.19}As$ top barrier, two monolayers of GaAs and finally an 8.7 nm layer of epitaxial Al. The top Al layer has been grown in the same chamber without breaking the vacuum. The two monolayers of GaAs are introduced to help passivate the wafer surface where the Al film is removed and to make the sample more compatible with the Al etchant, which does not attack GaAs. The two-dimensional electron gas (2DEG) is accumulated in the InAs quantum well.

## Device Fabrication

In the fabrication process, we have performed wet etching of the unnecessary epitaxial Aluminum with a type-D aluminum etchant to form JJs and SC loops. Then, we have grown a 30 nm-thick aluminum oxide layer through atomic layer deposition and fabricated a separate gate electrode on each junction with 5 nm-Ti and 20 nm-Au.

## Acknowledgments

Funding: This work was partially supported by a JSPS Grant-in-Aid for Scientific Research (S) (Grant No. JP19H05610); JST PRESTO (grant no. JPMJPR18L8); Advanced Technology Institute Research Grants; Ozawa-Yoshikawa Memorial Electronics Research Foundation; the US Department of Energy under Award No. DE-SC0019274.

## Author contributions

S.M. conceived the experiments. T.L., S.G., G.C.G., and M.J.M grew the quantum well wafer. S.M. fabricated and measured the device with T.I.. S.M., T.I., Y.S., and S.T. analyzed the data. S.N. and Y.T. gave the theoretical input. T.Y. performed the numerical calculation. S.T. supervised the study.

## Competing interest statement

The authors declare no competing interests.

## Supplementary Information

Supplementary notes 1-8 and supplementary figures S1-S8

# Figures

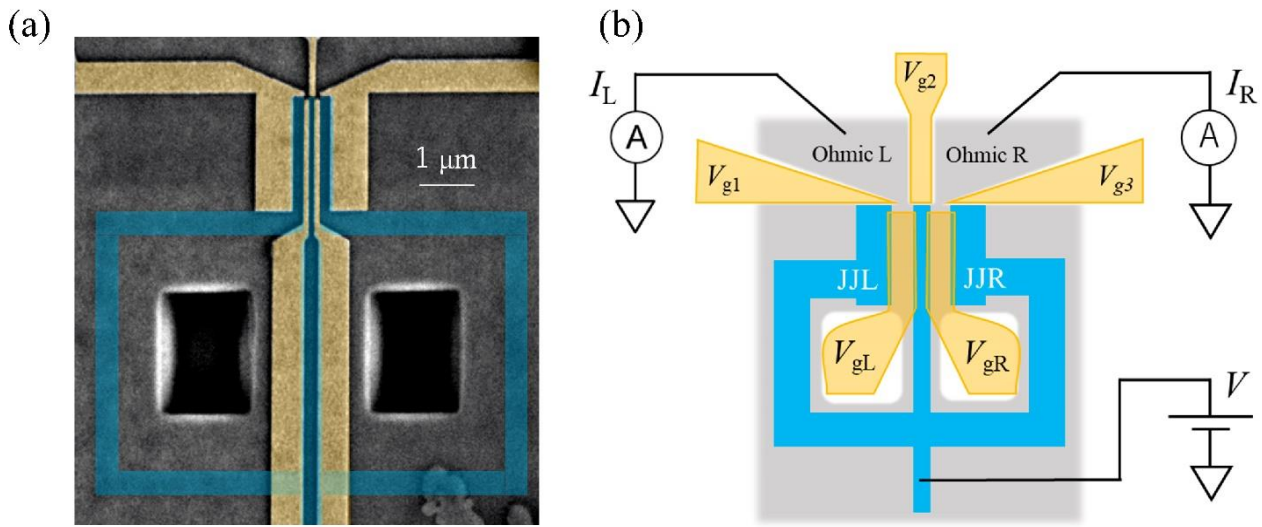

**Fig. 1 : A device for tunnel spectroscopy of the coupled JJs**
(a) A scanning electron microscopic image of the coupled planar JJ device. The blue and yellow regions represent the SC electrodes and the gate electrodes, respectively. Two JJs named JJL and JJR are coupled through a shared SC electrode and each is embedded in an SC loop. (b) A schematic image of the device. Tunnel currents through QPCL and QPCR are measured, and from the result, the differential conductance is calculated for each to implement the tunnel spectroscopy of JJL and JJR.

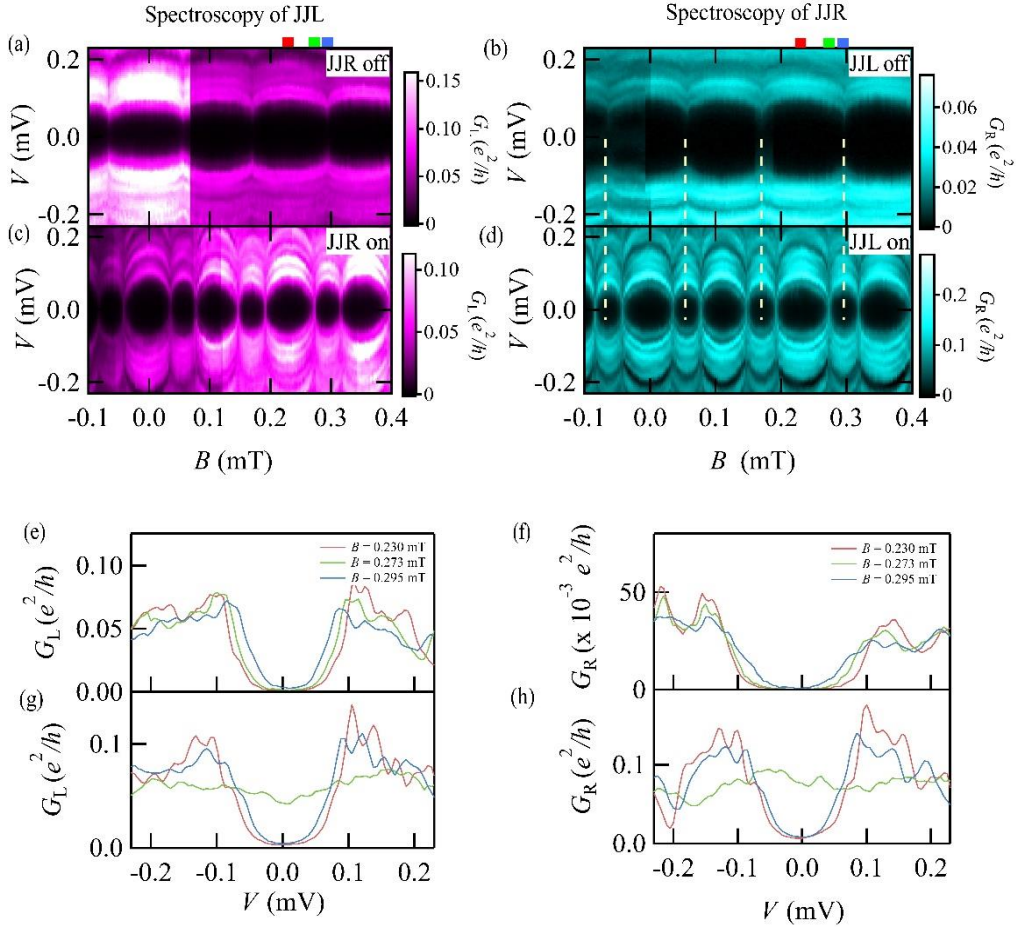

**Fig. 2 : Tunnel spectroscopy results of the single and coupled JJs**
(a) Tunnel spectroscopic result of the single JJL as a function of $B$. The SC gap oscillates to B, assigned to the feature of ABS expected in short JJs. (b) Tunnel spectroscopic result of the single JJR. Almost the same features as in the JJL result are found. The periodicity of the ABS oscillation is the same as that in (a) because the two loops hold the same area. The white dashed lines indicate the $B$ points giving the minimal SC gap. (c) Tunnel spectroscopic result of the JJL coupled with JJR. The result is drastically modulated from the single JJ case in (a). Especially, the SC gap becomes minimal away from $\phi_L = \phi_R = \pi$ where the gap becomes minimal in the single JJ cases. (d) Tunnel spectroscopic result of the JJR coupled with JJL. The same features observed in (c) were acquired. The consistency between (c) and (d) assures that the AMSs are constructed due to the coherent coupling of the two JJs. (e) $G_L$ vs. $V$ at $B$=0.230, 0.273, and 0.295 mT, respectively measured for the single JJL. (f) $G_R$ vs. $V$ measured for the single JJR. (g) $G_L$ vs. $V$ measured for the coupled JJL. The SC gap is closed at $B$= 0.273 mT while the other two curves hold the SC gap. (h) $G_R$ vs. $V$ measured for the coupled JJL. As with the coupled JJL case, the SC gap is closed at $B$= 0.273 mT while the other two curves hold the SC gap.

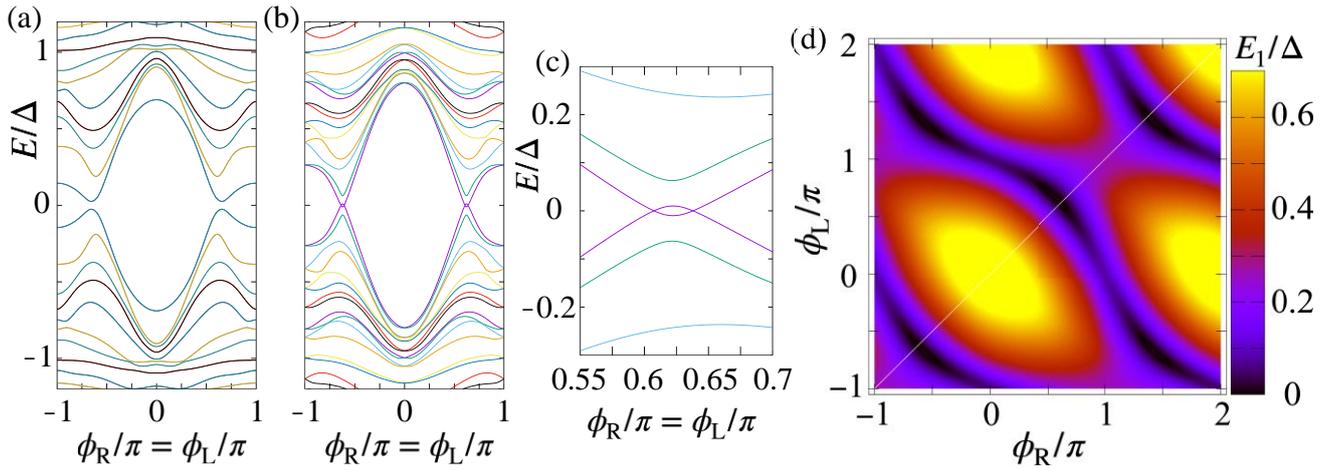

**Fig. 3 : The numerical calculation results indicating the AMS formation and SC gap closing**

AMS spectrum along $\phi_L = \phi_R$ without SOI (a) and with SOI (b). Owing to the SOI, the gap closing is observed. (c) Enlarged view of the gap closing in (b). (d) The energy of the positive lowest Andreev state $(E_1/\Delta)$ as a function of $\phi_L$ and $\phi_R$. The white line indicates the condition of $\phi_L = \phi_R$ corresponding to the experimental situation.

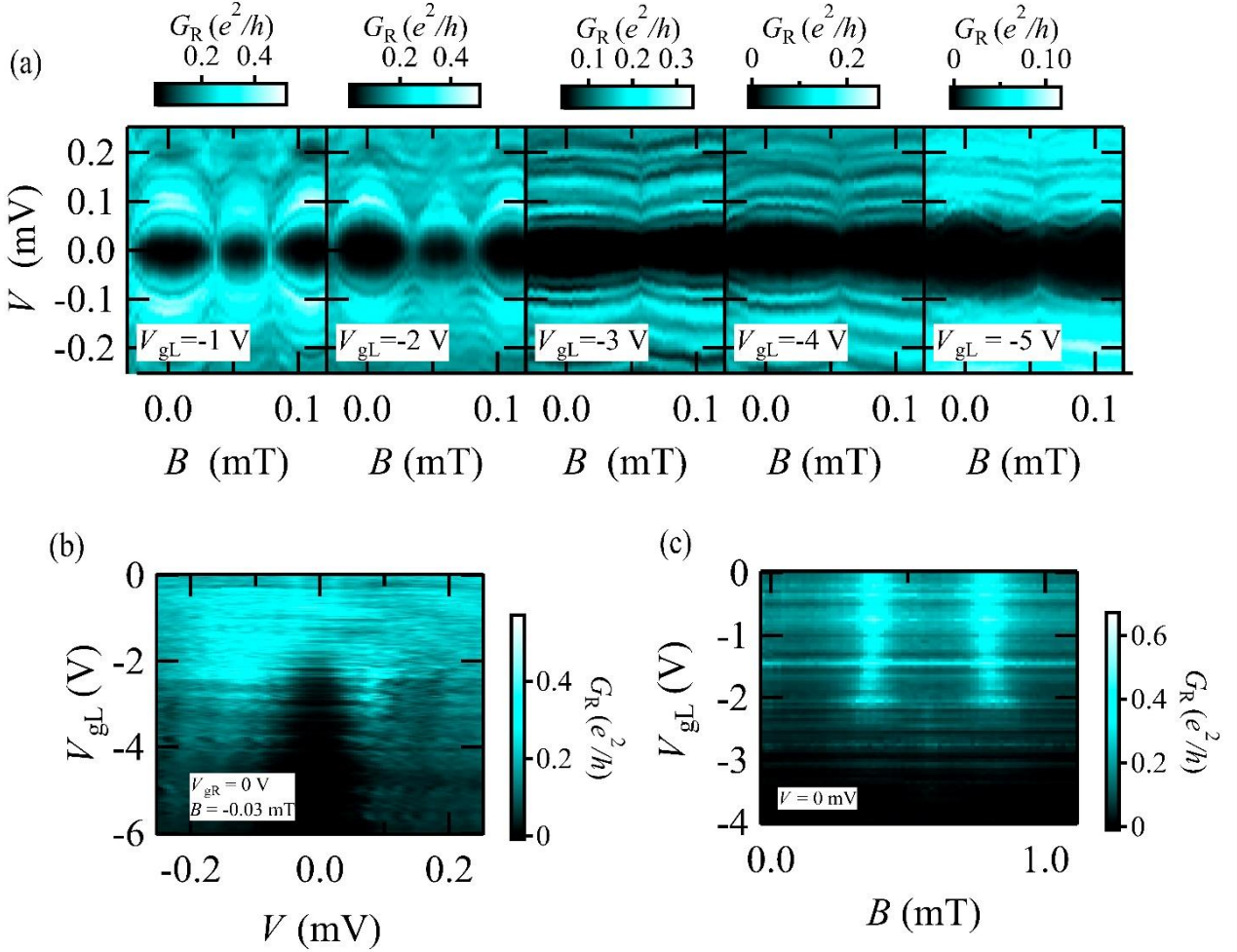

**Fig. 4 : Nonlocal gate control of the AMSs detected in the coupled JJR**
(a) $G_R$ as a function of $B$ and $V$ at several $V_{gL}$. The AMS features appear for $V_{gL} \geq -2$ V but disappear for $V_{gL} < -2$ V. This suggests that the JJL is pinched off at $V_{gL} \sim -2$ V. (b) $G_R$ as a function of $V$ and $V_{gL}$ at $V_{gR}= 0$ V and $B= -0.03$ mT. $B= -0.03$ mT corresponds to the SC gap closing point. Then the result indicates no gap structure for $V_{gL} \geq -2$ V while the SC gap grows as $V_{gL}$ becomes smaller than $-2$ V. This indicates that the SC gap closing is robust as long as the AMS is formed. (c) $G_R$ as a function of $B$ and $V_{gL}$ at $V=0$ mV. The two vertical bright lines represent $B$ producing the SC gap closing. The lines disappear around $V_{gL} \sim -2$ V, supporting that the SC gap closing is robust as long as the AMS is formed.